\documentclass{elsart}

\usepackage{graphicx}
\usepackage{amssymb}

\begin{document}

\begin{frontmatter}

\title{Fourier analysis of the flux-tube distribution in SU(3) lattice QCD}

\author{Arata~Yamamoto}
\ead{a-yamamoto@ruby.scphys.kyoto-u.ac.jp}
\address{Department of Physics, Faculty of Science, Kyoto University, \\Kitashirakawa, Sakyo, Kyoto 606-8502, Japan}

\begin{abstract}
This paper presents a novel analysis of the action/energy density distribution around a static quark-antiquark pair in SU(3) lattice quantum chromodynamics.
Using the Fourier transformation of the link variable, we remove the high-momentum gluon and extract the flux-tube component from the action/energy density.
When the high-momentum gluon is removed, the statistical fluctuation is drastically suppressed, and the singularities from the quark self-energy disappear.
The obtained flux-tube component is broadly distributed around the line connecting the quark and the antiquark. 
\end{abstract}

\end{frontmatter}

\section{Introduction}
Quark confinement is one of the most significant phenomena in quantum chromodynamics (QCD).
Due to the non-perturbative and non-Abelian nature of QCD, analytical derivation of quark confinement from the QCD Lagrangian has not yet been achieved.
While it has long been known that quark confinement is well described by
a ``string'' or ``flux tube'', its mechanism remains an unsolved problem \cite{Na74}.

Lattice QCD provides us with beneficial knowledge about quark confinement.
In lattice QCD, the formation of the flux tube is visualized by the analysis of the action/energy density distribution around a static quark-antiquark pair \cite{So87,Ha87,Gi90,Ba95,Ha96,Pe97,Pe98}.
At the positions of the quark and the antiquark, the action/energy
density is strongly enhanced due to the divergence of the quark self-energy.
In addition to such large perturbative contributions, a string-like structure is formed between the quark and the antiquark.
This is direct evidence of the flux-tube formation.

In recent works of lattice QCD, the static quark-antiquark potential
is analyzed by the Fourier transformation of the link variable \cite{Ya08,Ya09}.
These works clarify that quark confinement originates from an infrared gluon below 1.5 GeV in the Landau gauge.
When this infrared gluon is removed, the confinement potential completely disappears.
Conversely, when a high-momentum gluon above 1.5 GeV is removed, the short-range Coulomb potential disappears, and the quark-antiquark potential becomes a purely linear confinement potential.
Thus, by restricting the gluon field to the low-momentum region below 1.5 GeV, we can extract the essential contribution to color confinement.

In this letter, we apply this type of analysis by the Fourier transformation to the calculation of the action/energy density distribution.
Because the flux-tube contribution to the action/energy density is
rather smaller than the perturbative contribution, it is difficult to observe the flux-tube structure in traditional approaches.
For example, the endpoints of the flux tube are completely hidden by the perturbative singularities.
If the high-momentum gluon is removed, the unnecessary perturbative contribution would disappear while the flux-tube contribution would remain unchanged.
Therefore, by removing the high-momentum gluon, we can clearly observe the flux tube.
In other words, we can extract only the flux-tube component from the action/energy density distribution.

\section{Action density and energy density}

\begin{figure}[t]
\begin{center}
\includegraphics[scale=0.5]{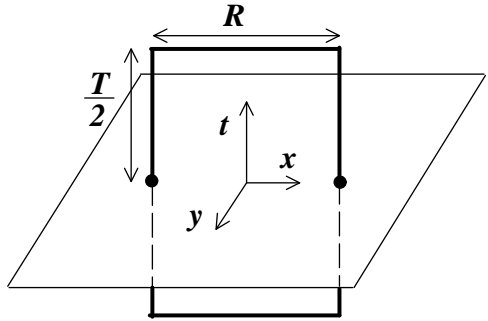}
\caption{\label{Fig1}
The Wilson loop $W(R,T)$ and the three-dimensional plane where the action density $\langle \rho (x) \rangle_W$ is measured.
The origin of the coordinate is placed in the center of the Wilson loop. 
}
\end{center}
\end{figure}

In lattice QCD, the action density is defined as
\begin{eqnarray}
\rho (x) = \beta \sum_{\mu>\nu} \left \{ 1-\frac{1}{N_c}{\rm ReTr } U_{\mu\nu}(x) \right \},
\label{rho}
\end{eqnarray}
where $U_{\mu\nu}(x)$ is the plaquette variable and $\beta=2N_c/g^2$.
The spatial distribution of the action density around a static quark-antiquark pair is obtained by measuring $\rho (x)$ around the Wilson loop at a certain time slice.
Its expectation value is given by
\begin{eqnarray}
\langle \rho (x) \rangle_W \equiv \frac{\langle \rho (x) W(R,T) \rangle}{\langle W(R,T) \rangle} - \langle \rho (x) \rangle,
\label{rhoW}
\end{eqnarray}
where $W(R,T)$ is the value of the Wilson loop with the size $R \times T$.
We consider only the case that the spatial size $R$ and the temporal
size $T$ are even numbers in the lattice unit.
The schematic figure is illustrated in Fig.~\ref{Fig1}.
The origin of the four-dimensional coordinate is placed in the center of the Wilson loop. 

The action density distribution $\langle \rho (x) \rangle_W$ is
independent of the time slice if the ground-state component is suitably dominated.
The temporal size $T$ should be sufficiently large to extract the ground-state distribution.
In the following, $\langle \rho (x) \rangle_W$ is measured at the central time slice of the Wilson loop ($t=0$).

The energy density $\varepsilon (x)$ is calculated identically to the action density by changing the relative sign between the spatial plaquettes and the temporal plaquettes.
Because these distributions around a static quark-antiquark pair are qualitatively similar except for the overall values, we mainly discuss the action density in the following.

\section{Momentum cutoff for the link variable}
By removing the high-momentum gluon, we investigated the behavior of the low-momentum gluon, which contributes to the confinement potential and the flux tube.
The lattice framework to remove momentum components of the
gluon field is introduced in recent works \cite{Ya08,Ya09}.
We used the three-dimensional version of this framework, i.e., we treated the spatial three-momentum of the gluon field.
Here, we only briefly explain the procedure.
The procedure is formulated as the following five steps.

Step 1. The SU(3) link variable $U_{\mu}(x)$ is generated by Monte Carlo simulation.
Here, the link variable must be fixed with a certain gauge.
In this letter, we choose the Landau gauge, which has a direct connection between the link variable and the gauge field.

Step 2. The momentum-space link variable ${\tilde U}_{\mu}(t,\vec{p})$ is given as
\begin{eqnarray}
{\tilde U}_{\mu}(t,\vec{p})=\frac{1}{L^3}\sum_{\vec{x}} U_{\mu}(x)\exp(i \vec{p} \cdot \vec{x}),
\end{eqnarray}
where $L$ is the number of lattice sites in the spatial direction.
The momentum space is also an $L^3$ lattice, and its lattice spacing is given by $a_p = 2\pi/La$.

Step 3. The high-momentum component of ${\tilde U}_{\mu}(t,\vec{p})$ is removed above the ultraviolet cutoff $\Lambda_{\rm UV}$.
The momentum-space link variable with the ultraviolet cutoff is defined as 
\begin{equation}
{\tilde U}_{\mu}^{\Lambda}(t,\vec{p})= \Bigg\{
\begin{array}{cc}
{\tilde U}_{\mu}(t,\vec{p}) & (|\vec{p}| \le \Lambda_{\rm UV})\\
0 & (|\vec{p}| > \Lambda_{\rm UV}).
\end{array}
\end{equation}

Step 4. The inverse Fourier transformation is performed as
\begin{eqnarray}
U'_{\mu}(x)=\sum_{\vec{p}} {\tilde U}_{\mu}^{\Lambda}(t,\vec{p})\exp(-i \vec{p} \cdot \vec{x}).
\end{eqnarray}
Note that $U'_{\mu}(x)$ is not an SU(3) matrix in general.
To obtain the SU(3) link variable, $U'_{\mu}(x)$ is projected onto an SU(3) element $U^{\Lambda}_{\mu}(x)$.
Such a projection is realized by maximizing the quantity
\begin{eqnarray}
{\rm ReTr}[ \{ U^{\Lambda}_{\mu}(x) \}^{\dagger} U'_{\mu}(x)].
\end{eqnarray}

Step 5. The expectation value of the operator $O$ is calculated from the link variable $U^{\Lambda}_{\mu}(x)$ instead of $U_{\mu}(x)$; i.e., $\langle O[U^\Lambda]\rangle$ instead of $\langle O[U]\rangle$.

Using this framework, we analyzed the low-momentum part of the action/energy density distribution.
This is one type of Fourier analysis of a spatial distribution.
However, unlike ordinary Fourier analysis, its Fourier component is not
that of the spatial distribution itself but rather of the gluon field.
In this sense, our analysis is a physical extraction rather than a numerical technique.

We comment on the gauge fixing in Step 1. 
In general, because the gauge transformation is nonlocal in momentum space, the momentum region of the gauge field is a gauge-dependent concept.
Therefore, the gauge fixing is necessary to remove a part of the momentum region.
In this letter, we show the numerical results of the Landau gauge.
It is noteworthy, however, that one can easily analyze another gauge in the same way.

It is worth noting that, as a by-product, our approach strongly suppresses the statistical noise of the gluon distribution.
Due to the large fluctuation of the ultraviolet gluon, precise observation of the flux tube is not easy in lattice QCD.
As shown in Sec.~5.3, the statistical noise is indeed suppressed by removing the ultraviolet gluon, and the gluon distribution becomes clear.

\section{Simulation setup}
We performed quenched lattice QCD simulation.
The gauge action is the SU(3) plaquette action with $\beta =6.0$.
The corresponding lattice spacing $a$ is about 0.10 fm, which is set so as to reproduce the string tension $\sigma=0.89$ GeV/fm.
The lattice volume is mainly $16^4$.
The configuration number is 500, and the statistical error is estimated by the jackknife method.
For statistical improvement, averages over the translationally invariant quantities are taken, if possible.
For example, $\langle \rho (x) W(R,T) \rangle$ is given by the convolution sum as
\begin{eqnarray}
\langle \rho (x) W(R,T) \rangle &=& \langle \frac{1}{L^3}\sum_{s} \rho (x+s) W(R,T,s) \rangle,
\end{eqnarray}
where $s$ is the position of the parallel-translated Wilson loop.

To enhance the ground-state component, the APE smearing method is applied to the spatial link variables of the Wilson loop \cite{Al87}.
In the calculation of the gluon distribution, it is necessary that the ground-state component is sufficiently dominant in the time slice where the gluon distribution is measured.
Due to the smearing method, the Wilson loop is rapidly dominated by the ground-state component.
In fact, the gluon distribution is found to be almost independent of $T$ in the range of $T \ge 4a$ in the present calculation.

\section{Numerical results}

\subsection{Quark-antiquark potential}

\begin{figure}[t]
\begin{center}
\includegraphics[scale=1.15]{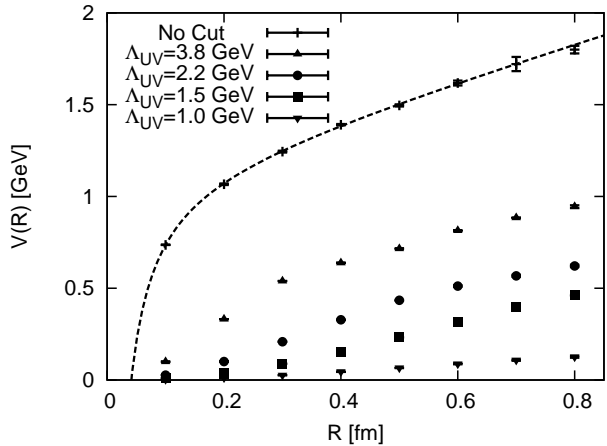}
\caption{\label{Fig2}
The quark-antiquark potential $V(R)$ with the ultraviolet cutoff $\Lambda_{\rm UV}$.
The ``No Cut'' is the original lattice QCD result.
}
\end{center}
\end{figure}

First, using the framework explained in Sec.~3, we demonstrate how the quark-antiquark potential is changed by the ultraviolet cutoff.
The original quark-antiquark potential is known to be expressed as
\begin{eqnarray}
V(R)=\sigma R -\frac{A}{R} +C,
\end{eqnarray}
where $R$ is the interquark distance.
The confinement potential is an infrared property.
On the other hand, the one-gluon-exchange Coulomb potential and the constant term, which originates mainly from the regularization for the short-range singularity, are ultraviolet properties.

In Fig.~\ref{Fig2}, the quark-antiquark potential with the ultraviolet cutoff $\Lambda_{\rm UV}$ is shown.
The result of the original lattice QCD is also shown in the figure.
When the high-momentum gluon is removed, the Coulomb potential and the constant term disappear because these terms are ultraviolet properties.
On the other hand, the confinement potential is almost unchanged up to $\Lambda_{\rm UV}=1.5$ GeV.
At $\Lambda_{\rm UV}=1.0$ GeV, the slope of the confinement potential suddenly decreases.
In Table \ref{tab1}, we list the asymptotic string tension $\sigma_{\rm asym}$, which is estimated by fitting the quark-antiquark potential in $ R \ge 0.4$ fm with a linear function $\sigma_{\rm asym}R + {\rm const}$.
In $\Lambda_{\rm UV}\ge 1.5$ GeV, the asymptotic string tension is
almost equal to the original value.

This behavior of the quark-antiquark potential is consistent with the case of the four-dimensional cutoff \cite{Ya08,Ya09}.
In the Landau gauge, the low-momentum gluon below 1.5 GeV is relevant
for quark confinement from both the three-dimensional and the four-dimensional viewpoints.

\begin{table}[b]
\newcommand{\m}{\hphantom{$-$}}
\newcommand{\cc}[1]{\multicolumn{1}{c}{#1}}
\renewcommand{\tabcolsep}{1pc} 
\renewcommand{\arraystretch}{1} 
\caption{\label{tab1}
The ultraviolet cutoff $\Lambda_{\rm UV}$, the asymptotic string tension $\sigma_{\rm asym}$ of the quark-antiquark potential, and the vacuum action density $\langle \rho (x) \rangle$.
The statistical error of $\langle \rho (x) \rangle$ is omitted because it is negligibly small.
}
\begin{center}
\begin{tabular}{ccccc}
\hline\hline
$\Lambda_{\rm UV}$ [GeV] &$\sigma_{\rm asym}$ [GeV/fm] & $\langle \rho (x) \rangle$ [$a^{-4}$] \\
\hline
No Cut & 0.89 & 14.51 \\
3.8 & 0.824(31) & 2.57 \\
2.2 & 0.801(67)& 0.60 \\
1.5 & 0.799(18) & 0.20\\
1.0 & 0.208(4) & 9.5$\times 10^{-3}$\\
\hline\hline
\end{tabular}
\end{center}
\end{table}

\begin{figure*}[t]
\begin{center}
\includegraphics[scale=1.5]{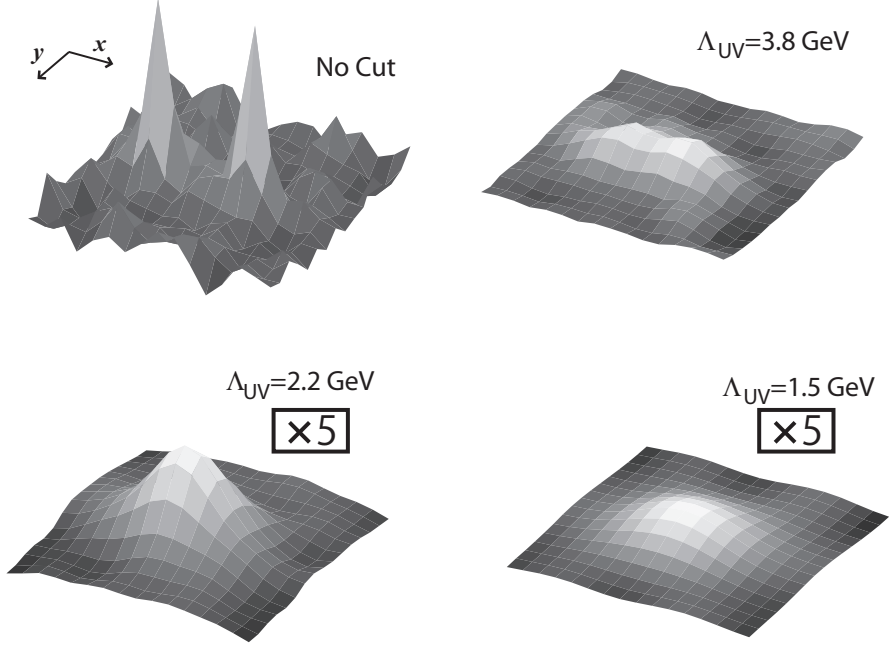}
\caption{\label{Fig3}
The action density distribution $\langle \rho (x) \rangle_W$ around a static quark-antiquark pair.
The separation between the quark and antiquark is $R=0.6$ fm.
``$\times 5$''s mean that the lower figures are five times enlarged in the vertical direction compared to the upper figures.
The ``No Cut'' (the upper left) is the original lattice QCD result without the ultraviolet cutoff, and its statistical error is relatively large.
}
\end{center}
\end{figure*}

\subsection{Action density in vacuum}
Before the action density distribution with a static quark-antiquark pair, we calculated the action density without color sources, i.e., in vacuum.
The vacuum action density relates to the gluon condensate in continuum QCD, which produces the trace anomaly \cite{Ba81,Gi81}.
In the naive continuum limit, the action density corresponds to the gluon condensate; however, at finite lattice spacing, it is dominated by perturbative corrections rather than the non-perturbative gluon condensate.
In short, a large part of the vacuum action density on the lattice is perturbatively generated.

The vacuum action density $\langle \rho (x) \rangle$ with the ultraviolet cutoff is listed in Table \ref{tab1}.
In the table, the statistical error is omitted because it is negligibly small.
When the high-momentum gluon is removed, the vacuum action density drastically decreases.
For example, at $\Lambda_{\rm UV}= 1.5$ GeV, the vacuum action density is reduced to about 1 \% compared to the original value.
The remaining small component would lead to non-perturbative properties of QCD vacuum.

\subsection{Action density with a quark-antiquark pair}

The action density distribution around a static quark-antiquark pair is measured by Eq.~(\ref{rhoW}).
The ultraviolet cutoff is introduced to the action density as
\begin{eqnarray}
\label{eq9}
\langle \rho [U^\Lambda] \rangle_W \equiv \frac{\langle \rho [U^\Lambda] W[U] \rangle}{\langle W[U] \rangle} - \langle \rho [U^\Lambda] \rangle.
\end{eqnarray}
The arguments, such as $x$, are abbreviated for simplicity.
The physical interpretation is the spatial distribution of the low-momentum gluon around a physical quark-antiquark pair.
As another choice, one can introduce the ultraviolet cutoff not only for $\rho [U]$ but also for $W[U]$, and its result is expected to be qualitatively similar to that of Eq.~(\ref{eq9}).

In Fig.~\ref{Fig3}, we display the action density distribution $\langle \rho (x) \rangle_W$ with the ultraviolet cutoff $\Lambda_{\rm UV}=1.5$ GeV, 2.2 GeV, and 3.8 GeV.
The interquark distance between the quark and the antiquark is $R=0.6$ fm.
The overall sign of $\langle \rho (x) \rangle_W$ is flipped in the figure, which is only a matter of definition.
Because the absolute value of $\langle \rho (x) \rangle_W$ at
$\Lambda_{\rm UV}=1.5$ GeV and 2.2 GeV is small, these data in
Fig.~\ref{Fig3} are enlarged by a factor of five compared to the other ones.
The action density distribution in original lattice QCD without the cutoff is also displayed (``No Cut'' in the upper left); however, its statistical error is relatively large.
In original lattice QCD, the action density is strongly enhanced at the positions of the quark and the antiquark.
In contrast, the flux-tube structure is difficult to observe due to such singular peaks and the large statistical fluctuation.
Both the singular peaks and the large fluctuation originate from the perturbative property of the action density.
When the high-momentum gluon above 3.8 GeV is removed (the upper right), these perturbative contributions are drastically suppressed, and the flux-tube structure connecting the quark and the antiquark becomes clear.
At $\Lambda_{\rm UV}=1.5$ GeV and 2.2 GeV (the lower right and the lower left, respectively), the two peaks  seem to disappear, and the action density is distributed around the origin.

Apart from the vacuum contribution, which is translationally invariant, the action density distribution at $\Lambda_{\rm UV}=1.5$ GeV is broadly distributed around the midpoint between the quark and the antiquark.
In the calculation of the quark-antiquark potential, this low-momentum gluon leads to the linear confinement potential over the entire range of $R$.
Therefore, it is considered that this action density distribution corresponds to the confining flux tube.

\begin{figure}[t]
\begin{center}
\includegraphics[scale=1.15]{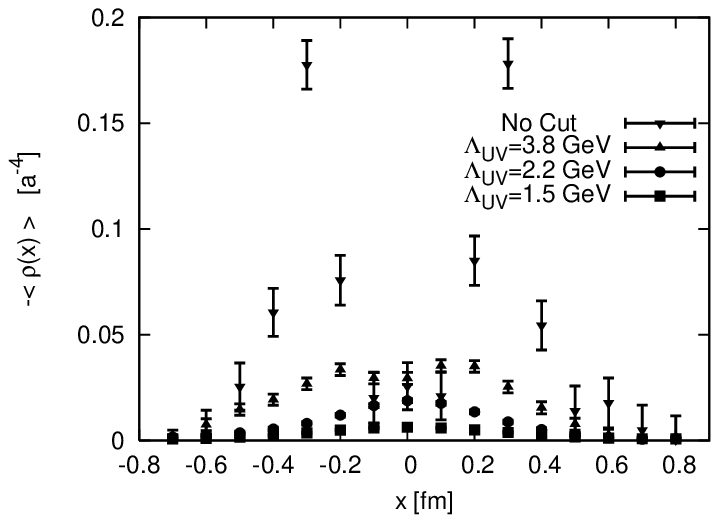}
\caption{\label{Fig4}
The action density distribution $\langle \rho (x) \rangle_W$ along the $x$-axis.
The interquark distance $R$ is 0.6 fm. 
The quark and the antiquark are located on $x=0.3$ fm and $x=-0.3$ fm, respectively.
The ``No Cut'' is the original lattice QCD result.
}

\includegraphics[scale=1.15]{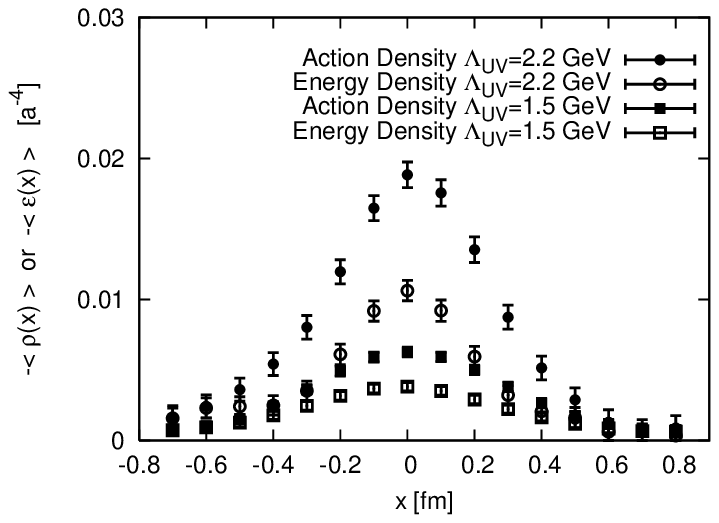}
\caption{\label{Fig5}
The action density distribution $\langle \rho (x) \rangle_W$ and the energy density distribution $\langle \varepsilon (x) \rangle_W$ along the $x$-axis.
The interquark distance $R$ is 0.6 fm. The quark and the antiquark are located on $x=0.3$ fm and $x=-0.3$ fm, respectively.
}
\end{center}
\end{figure}

In Fig.~\ref{Fig4}, we plot the action density distribution $\langle \rho (x) \rangle_W$ along the $x$-axis, i.e., in the longitudinal direction of the quark-antiquark separation.
In original lattice QCD without the ultraviolet cutoff (``No Cut''), the action density has two self-energy peaks at the positions of the quark and the antiquark, $x=0.3$ fm and $x=-0.3$ fm, respectively.
When the ultraviolet cutoff is introduced, these self-energy peaks are drastically suppressed.
Moreover, the absolute values of the action density and the statistical fluctuation become small, as in the case of the vacuum action density.
The results at $\Lambda_{\rm UV}=1.5$ GeV and 2.2 GeV are also shown in Fig.~\ref{Fig5}.
The action density distribution has a maximum at the origin, and the self-energy peaks seem to disappear.
Although the self-energy peaks would also include a nonperturbative contribution, it is too small to distinguish from the flux tube.
The endpoints of the flux tube are not sharp, spreading outside the positions of the quark and the antiquark.
In the case of $R=0.6$ fm, the longitudinal shape of the flux tube resembles a broad mountain rather than a plateau.

We also show the energy density distribution $\langle \varepsilon (x) \rangle_W$ along the $x$-axis in Fig.~\ref{Fig5}.
The absolute value of the energy density is smaller than that of the action density due to the cancellation between the chromoelectric contribution and the chromomagnetic contribution.
Apart from the absolute value, the overall shape of the energy density distribution is similar to that of the action density distribution.

\begin{figure}[t]
\begin{center}
\includegraphics[scale=1.15]{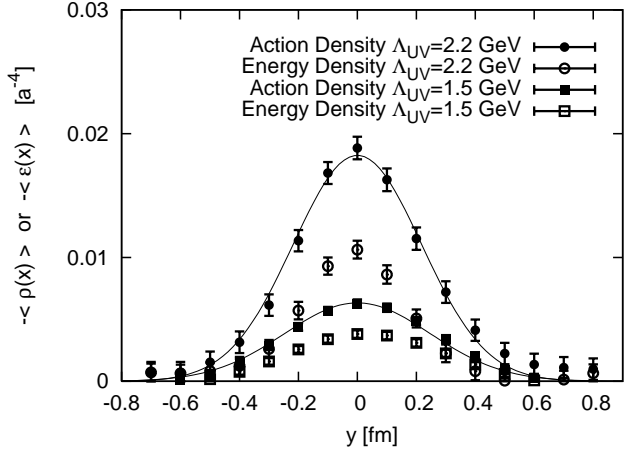}
\caption{\label{Fig6}
The action density distribution $\langle \rho (x) \rangle_W$ and the energy density distribution $\langle \varepsilon (x) \rangle_W$ along the $y$-axis.
The interquark distance $R$ is 0.6 fm. 
The solid lines are the results of fitting to a the Gaussian function $\rho_0 \exp (-y^2/\delta^2)$.
}
\end{center}
\end{figure}

To estimate the width of the flux tube, we fit the action density
distribution along the $y$-axis to a the Gaussian form, $\rho_0 \exp (-y^2/\delta^2)$.
It is seen that the Gaussian form can well reproduce the lattice data, as shown in Fig.~\ref{Fig6}.
The best-fit width parameter $\delta$ is 0.31$\pm$0.01 fm at $\Lambda_{\rm UV}=2.2$ GeV, and 0.35$\pm$0.01 fm at $\Lambda_{\rm UV}=1.5$ GeV.
These values are comparable to the flux-tube width of earlier works in
the standard lattice QCD \cite{Gi90,Ba95,Ha96}.

\begin{figure}[t]
\begin{center}
\includegraphics[scale=1.4]{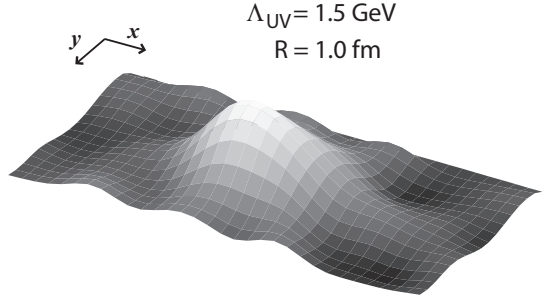}
\caption{\label{Fig7}
The action density distribution $\langle \rho (x) \rangle_W$ for the interquark distance $R=1.0$ fm.
This calculation is performed on a 32$\times 16^3$ lattice.
}
\end{center}
\end{figure}

Next, we analyze how the flux-tube shape depends on the interquark distance $R$.
As shown in Fig.~\ref{Fig3}, the longitudinal length and the transverse width are almost the same at $R=0.6$ fm and $\Lambda_{\rm UV}=1.5$ GeV.
The overall shape seems to be isotropic and far from a ``string'' or ``tube''.
This is because the transverse width of the flux tube is fairly large.
To approach the gluon distribution to the tube-like shape, the interquark distance must be enlarged \cite{Ba95,Ha96}.
Here, we use a 32($x$-axis)$\times 16^3$ lattice instead of a $16^4$ lattice and extend the interquark distance to $R=1.0$ fm.
As shown in Fig.~\ref{Fig7}, the distribution is stretched in the longitudinal direction, and the longitudinal length becomes larger than the transverse width.
Compared to the result of $R=0.6$ fm, the flux tube at $R=1.0$ fm approaches a broad tube-like shape.

\section{Summary}
Using the Fourier transformation of the link variable in SU(3) lattice QCD, we have extracted the flux-tube component from the action/energy density distribution.
By removing the high-momentum gluon above 1.5 GeV in the Landau gauge, we can eliminate the unnecessary perturbative contribution, such as the singular peaks at the positions of the quark and the antiquark.
As a by-product, the ultraviolet statistical fluctuation is also drastically suppressed.

The resultant flux-tube component is broadly, almost isotropically, distributed when the interquark distance is not large, as shown in Fig.~\ref{Fig3}.
When the interquark distance becomes larger, the flux-tube component is stretched in the longitudinal direction and approaches a broad tube-like shape, as shown in Fig.~\ref{Fig7}.
These distributions are the essential shapes of the confining flux tube.

Finally, we comment on the L\"{u}scher term, which is a $-\pi /(12R)$ correction to the quark-antiquark potential due to the string fluctuation \cite{Lu80,Lu81}.
Its functional form is similar to the perturbative Coulomb potential, but its origin is nonperturbative.
While the short-range Coulomb potential vanishes due to the ultraviolet
cutoff, as shown in Sec.~5.1, it is nontrivial whether or not the long-range L\"{u}scher term remains.
It is interesting to analyze the L\"{u}scher term in this framework.

\section*{Acknowledgments}
The author is very grateful to H.~Suganuma for many beneficial discussions.
This work is supported by a Grant-in-Aid for Scientific Research [(C) No.~20$\cdot$363] in Japan and by the Global COE Program, ``The Next Generation of Physics, Spun from Universality and Emergence''.
The lattice QCD calculations are done on NEC SX-8R at Osaka University.

\end{document}